\documentclass[sigconf, nonacm]{acmart}

\usepackage{threeparttable}
\usepackage{multirow}
\usepackage{subfigure}
\usepackage{caption}
\usepackage{float}
\usepackage{soul}
\usepackage{balance}
\usepackage{ulem}
\usepackage{tablefootnote}
\newcommand{\tabincell}[2]{\begin{tabular}{@{}#1@{}}#2\end{tabular}}
\newtheorem{definition}{Definition}

\newcommand\vldbdoi{10.14778/3514061.3514074}
\newcommand\vldbpages{XXX-XXX}
\newcommand\vldbvolume{15}
\newcommand\vldbissue{6}
\newcommand\vldbyear{2022}
\newcommand\vldbauthors{\authors}
\newcommand\vldbtitle{\shorttitle} 
\newcommand\vldbavailabilityurl{}

\newcommand\vldbpagestyle{empty}

\begin{document}
\pagenumbering{arabic}          
\pagestyle{plain}

\title{PGE: Robust Product Graph Embedding Learning for Error Detection}







\author{Kewei Cheng}
\email{viviancheng@cs.ucla.edu}
\affiliation{\institution{University of California, Los Angeles}}

\author{Xian Li}
\email{xianlee@amazon.com}
\affiliation{
Amazon.com 
}

\author{Yifan Ethan Xu}
\email{xuyifa@amazon.com}
\affiliation{
Amazon.com 
}

\author{Xin Luna Dong}
\email{lunadong@gmail.com}
\affiliation{
Facebook.com 
}

\author{Yizhou Sun}
\email{yzsun@cs.ucla.edu}
\affiliation{\institution{University of California, Los Angeles}}

\begin{abstract}
Although product graphs (PGs) have gained increasing attentions in recent years for their successful applications in product search and recommendations, the extensive power of PGs can be limited by the inevitable involvement of various kinds of errors. Thus, it is critical to validate the correctness of triples in PGs to improve their reliability. Knowledge graph (KG) embedding methods have strong error detection abilities. Yet, existing KG embedding methods may not be directly applicable to a PG due to its distinct characteristics: (1) PG contains rich textual signals, which necessitates a joint exploration of both text information and graph structure; (2) PG contains a large number of attribute triples, in which attribute values are represented by free texts. Since free texts are too flexible to define entities in KGs, traditional way to map entities to their embeddings using ids is no longer appropriate for attribute value representation; (3) Noisy triples in a PG mislead the embedding learning and significantly hurt the performance of error detection. To address the aforementioned challenges, we propose an end-to-end noise-tolerant embedding learning framework, PGE, to jointly leverage both text information and graph structure in PG to learn embeddings for error detection.
Experimental results on real-world product graph demonstrate the effectiveness of the proposed framework comparing with the state-of-the-art approaches.
\end{abstract}

\maketitle

\pagestyle{\vldbpagestyle}
\begingroup\small\noindent\raggedright\textbf{PVLDB Reference Format:}\\
\vldbauthors. \vldbtitle. PVLDB, \vldbvolume(\vldbissue): \vldbpages, \vldbyear.\\
\href{https://doi.org/\vldbdoi}{doi:\vldbdoi}
\endgroup
\begingroup
\renewcommand\thefootnote{}\footnote{\noindent
This work is licensed under the Creative Commons BY-NC-ND 4.0 International License. Visit \url{https://creativecommons.org/licenses/by-nc-nd/4.0/} to view a copy of this license. For any use beyond those covered by this license, obtain permission by emailing \href{mailto:info@vldb.org}{info@vldb.org}. Copyright is held by the owner/author(s). Publication rights licensed to the VLDB Endowment. \\
\raggedright Proceedings of the VLDB Endowment, Vol. \vldbvolume, No. \vldbissue\ %
ISSN 2150-8097. \\
\href{https://doi.org/\vldbdoi}{doi:\vldbdoi} \\
}\addtocounter{footnote}{-1}\endgroup

\ifdefempty{\vldbavailabilityurl}{}{
\vspace{.3cm}
\begingroup\small\noindent\raggedright\textbf{PVLDB Artifact Availability:}\\
The source code, data, and/or other artifacts have been made available at \url{\vldbavailabilityurl}.
\endgroup
}

\section{Introduction}\label{sec:intro}
With the rapid growth of the internet, e-commerce websites such as Amazon, eBay, and Walmart provide important channels
to facilitate online shopping and business transactions. As an effective way to organize product-related information, product knowledge graphs (PGs)~\cite{dong2020autoknow} have attracted increasing attentions in recent years 
by empowering many real-world applications, such as product search and recommendations~\cite{anelli2018knowledge}. 

\begin{figure}
\centering
\begin{minipage}[t]{\linewidth}
\centering       
        \resizebox{\linewidth}{!}{
        \begin{threeparttable} 
        \begin{tabular}{cccccc}
        \toprule
        Title & Category & Flavor & Ingredient &   \\
        \midrule
        \tabincell{c}{Brand A Tortilla Chips \\ Spicy Queso, \\6 - 2 oz bags}\tnote{$\dag$} & \tabincell{c}{chips-and\\-crisps} & Spicy Queso &  \tabincell{c}{Ground Corn, Chipotle \\ Pepper Powder,\\ Paprika Extract, Spices}  \\
        \midrule
        \tabincell{c}{Brand B Bean Chips\\ \underline{Spicy Queso}, \\High Protein and Fiber, \\ Gluten Free, Vegan Snack, \\ 5.5 Ounce (Pack of 6)}\tnote{$\dag$} &
        \tabincell{c}{chips-and\\-crisps} & \underline{Cheddar}  &  \tabincell{c}{Navy Beans,\\ Cayenne Pepper,\\ Paprika Extract,\\ Dehydrated Spices} \\
        \midrule
        \tabincell{c}{Carolina Reaper Spicy \\ Peanut Brittle} & \tabincell{c}{candy\\-brittle} & \tabincell{c}{Carolina \\ Reaper Spicy} & \tabincell{c}{Peanuts, Sugar, \\ Carolina Reaper} \\
        \bottomrule 
        \end{tabular}
\begin{tablenotes} 
\footnotesize
\item[$\dag$] We mask the brand of the products to avoid revealing sensitive information.
\end{tablenotes} 
\end{threeparttable}
}
\end{minipage}%
\! 
\begin{minipage}[t]{\linewidth}
\centering       
\includegraphics[width=\linewidth]{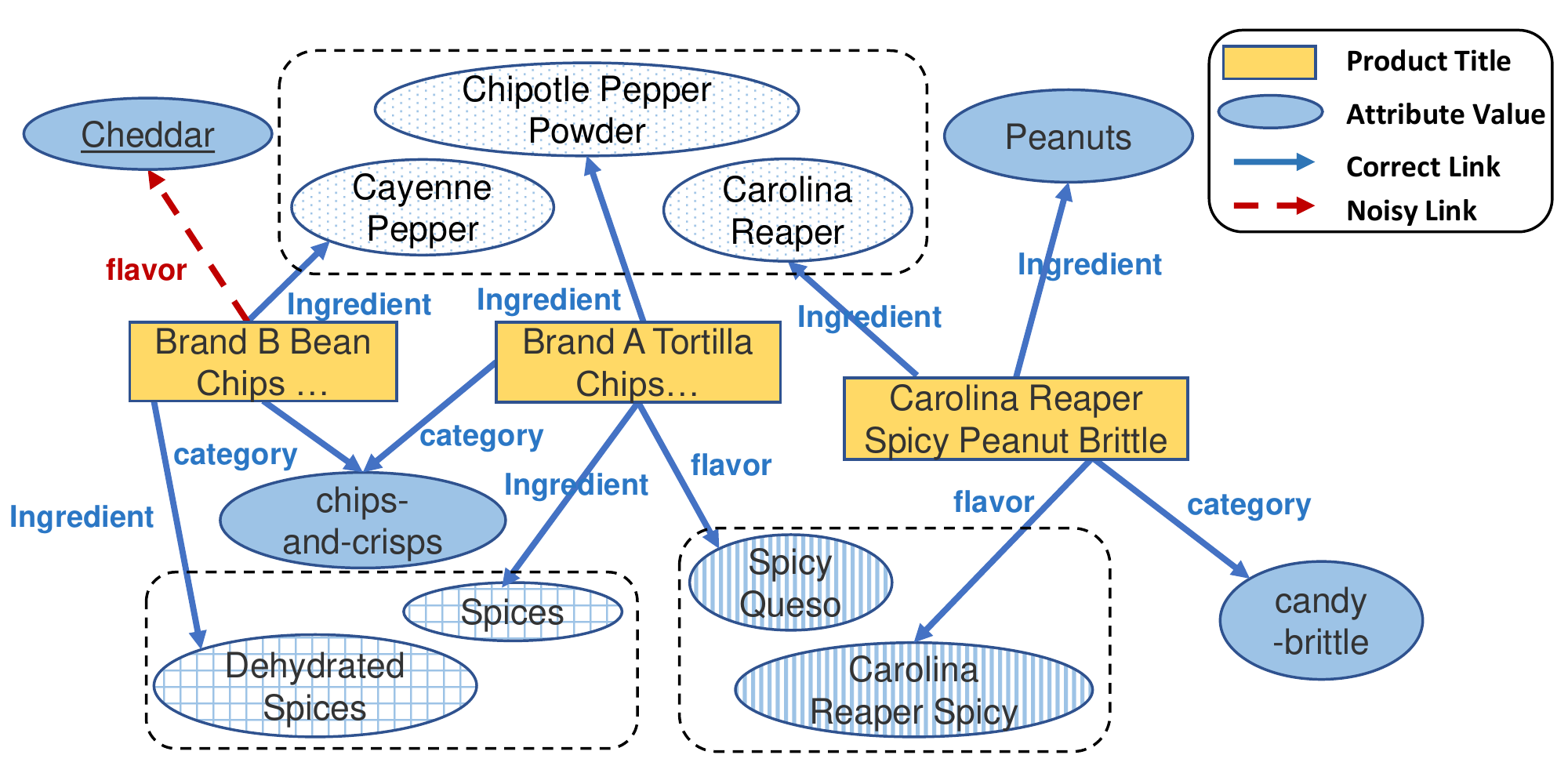}
\end{minipage}%
\captionof{figure}{\small{An example PG and its corresponding product catalog data. We underline the incorrect attribute value in the table whose ground truth value is given in its product title. Attribute values with similar semantic meanings are filled with the same pattern and gathering together with a dotted frame.}
}\label{fig::PG}
\vspace{-0.1in}
\end{figure}

A PG is a knowledge graph (KG) that describes product attribute values. It is constructed based on product catalog data (Fig.~\ref{fig::PG} shows an example). In a PG, each product is associated with multiple attributes
such as product brand, product category, and other information related to product properties such as flavor and ingredient. Different from traditional KGs, where most triples are in the form of (head entity, relation, tail entity),
the majority of the triples in a PG have the form of (product, attribute, attribute value), where the attribute value is a short text, e.g., \textit{(``Brand A Tortilla Chips Spicy Queso, 6 - 2 oz bags'', flavor, ``Spicy Queso'')}. We call such triples \textit{attribute triples}. 

\begin{figure}
\centering
\includegraphics[width=0.8\linewidth]{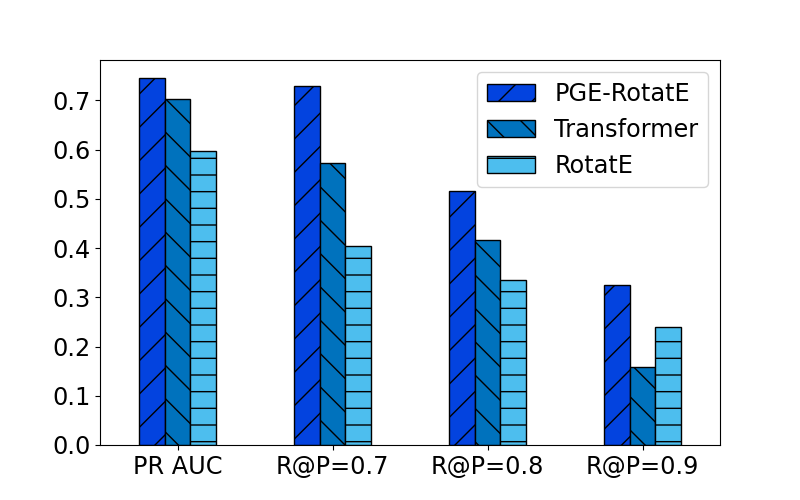}
\caption{\small{PGE improves over KG embedding method RotatE by 24.7\% and transformer by 4\% on PR AUC in transductive setting. It also shows significant improvement on R@P metric. R@P = 0.7 shows the recall when the precision is 0.7, etc.}}\label{fig:sample_exp}
\vspace{-0.1in}
\end{figure}

A vast majority of the product catalog data are provided by individual retailers. These self-reported data inevitably contain many kinds of errors, including conflicting, erroneous, and ambiguous values.
When such errors are ingested by a PG, they lead to unsatisfying performance of its downstream applications.
Due to the huge volume of products in a PG, manual validation is not feasible. An automatic validation method is in urgent need.

Knowledge graph embedding (KGE) methods currently hold the state-of-the-art in learning effective representations for multi-relational graph data. It aims to learn the network structure which triples should comply. KG embedding methods have shown promising performances in error detection (i.e., determine whether a triple is correct or not) in KGs~\cite{an2018accurate,xu2016knowledge}. For example, the PG structure in Fig.~\ref{fig::PG} indicates a strong correlation between the ingredient ``pepper'' and flavor ``spicy'' because they are connected through multiple products. By verifying its consistency with the network structure, the errors like
\textit{(``Brand B Bean Chips \underline{Spicy Queso}, High Protein and Fiber, Gluten Free, Vegan Snack, 5.5 Ounce (Pack of 6)'', flavor, ``\underline{Cheddar}'')} can be easily identified. 
Unfortunately, the existing KG embedding methods cannot be directly used to detect errors in a PG because of the following challenges.

\textbf{C1: PG contains rich textual information.} Products in a PG are often described by short texts like their titles and descriptions that contain rich information about their attributes.
For example, the product title ``Brand A Tortilla Chips Spicy Queso, 6 - 2 oz bags'' covers multiple attributes, including brand, product category, flavor, and size. 
We can easily verify the correctness of these attributes against the product title. 
In addition, the attribute values in PG are free texts. Thus the traditional way of mapping entity ids to their embeddings is no longer appropriate.
As shown in Fig.~{\ref{fig::PG}}, when the attribute values ``Chipotle Pepper Powder'' and ``Carolina Reaper'' (a kind of pepper) are modeled as two independent entities using their ids, the strong conceptual correlation between the ingredient ``pepper`` and the flavor ``spicy`` is lost.
Although several recent publications~\mbox{\cite{xie2016representation,xiao2017ssp}} tried to exploit the rich textual information in KGs, the network structure and text information were not jointly encoded into a unified representation. For example, text-based representation and structure-based representation were learned by separate loss functions and integrated into one joint representation by a linear combination~\mbox{\cite{xie2016representation,an2018accurate}}.

\textbf{C2: PG contains a large number of unseen attribute values.} 
The flexibility of textual attribute values also makes handling ``unseen attribute values'' challenging. In the example as shown in Fig.~{\ref{fig::PG}}, we can learn the representation of ``Chipotle Pepper Powder'' during training, but an unobserved attribute value with similar semantic meaning, such as ``Chipotle Pepper'' might be given for validation.
Conventional KG embedding models cannot deal with this inductive setting because they have no representations for the entities outside of KGs.

\textbf{C3: Existing noisy data in PG make it hard to learn a reliable embedding model.}  Getting a reliable embedding model for error detection in a PG requires clean data for training. However, noise widely existing in a PG can mislead the embedding model to learn the wrong structure information, which may severely downgrade its performance in error detection. 
\begin{table}
    \centering 
    \footnotesize
        \begin{tabular}{cccc}
        \toprule
        Methods & \tabincell{c}{Modeling \\ graph structure} & \tabincell{c}{Modeling \\ textual data} &  Noise-aware  \\
        \midrule
        \tabincell{c}{Structure based \\ KG embedding~\cite{ bordes2013translating,yang2014embedding,trouillon2016complex,sun2019rotate}} & \checkmark  &  &  \\
        \midrule
        \tabincell{c}{Text and KG \\ joint embedding~\cite{xie2016representation,xiao2017ssp, an2018accurate}} & \checkmark & \checkmark&  \\
        \midrule
        \tabincell{c}{Noise-aware \\ KG embedding~\cite{xie2018does}} & \checkmark  & & \checkmark \\
        \midrule
        PGE & \checkmark & \checkmark & \checkmark \\
        \bottomrule 
        \end{tabular}
    \captionof{table}{\small{Capabilities of different methods.}}\label{tab::Capabilities}
\vspace{-0.1in}
\end{table}

No existing approach is capable of tackling all aforementioned challenges, as shown in Table~\ref{tab::Capabilities}. Therefore, in this paper, we aim to answer this challenging research question: \textit{how to generate embeddings for a text-rich, error-prone knowledge graph to facilitate error detection?} We present a novel embedding learning framework, robust \textbf{P}roduct \textbf{G}raph \textbf{E}mbedding (\textbf{PGE}), to learn effective embeddings for such knowledge graphs. There are two key underlying ideas for our framework. First, our embeddings seamlessly combine the signals from the textual information of attribute triples, and the structural information in the knowledge graph. We do this by applying a CNN encoder to learn text-based representations for product titles and attribute values, and then integrating these text-based representations into the triplet structure to capture the underlying patterns in the knowledge graph. Second, we present a noise-aware loss function to prevent noisy triples in the PG from misguiding the embeddings during training. For each positive instance in the training data, our model predicts the correctness of the triple according to its consistency with the rest of the triples in the KG, and downweights an instance when the confidence of its correctness is low. As shown in Table~\ref{tab::Capabilities}, PGE is able to model both textual evidence and graph structure, and is robust to noise. 

Our proposed model is generic and scalable. First, it applies not only on the product domain, but also excel in other domains such as on Freebase KG, as we show in our experiments. Second, through careful choices of the deep learning models, our model can be trained on KGs with millions of nodes within a few hours, and are robust to noises and unseen values that are inherent in real data. In summary, this paper makes the following contributions.

\begin{itemize}


\item We propose an end-to-end noise-tolerant embedding learning framework, PGE, to jointly leverage both text information and graph structure in PG to learn embeddings for error detection.

\item We propose a novel noise-aware mechanism to incorporate triple confidence into PGE model to detect noise while learning knowledge representations simultaneously.

\item We evaluate PGE on a real-world PG w. millions of nodes generated from public Amazon website and show that we are able to improve over state-of-the-art methods on average by 18\% on PR AUC in transductive setting as summarized in Figure~\ref{fig:sample_exp}.
\end{itemize}

\section{Preliminaries and Problem Definition}\label{sec:prob_def}

We first formally define two important concepts: attribute triples and product graph.
\begin{definition}
Attribute triples
\label{def:triple}
\end{definition}
An attribute triple can be represented as $(t, a, v)$, where
its subject entity $t$ is a product sold on Amazon (e.g., a product with title ``Brand A Tortilla Chips Spicy Queso, 6 - 2 oz bags''), its object entity $v$ is an attribute value (e.g., ``spicy queso''), and $a$ is an attribute to connect $t$ and $v$ (e.g., flavor). Both $t$ and $v$ are represented as unstructured short texts. An attribute triple $(t, a, v)$ is {\em incorrect} if its attribute value $v$ does not correctly describe the product $t$. For example, \textit{(``Brand B Bean Chips \underline{Spicy Queso}, High Protein and Fiber, Gluten Free, Vegan Snack, 5.5 Ounce (Pack of 6)'', flavor, ``\underline{Cheddar}'')}
in Fig.~\ref{fig::PG} is an incorrect attribute triple.
\begin{definition}
Product Graph
\end{definition}
A 
Product graph (PG) is a KG that describes product attribute values. Formally, we represent a product graph as $\mathcal G = \{T, A, V, O \}$, where $T$ is a set of product titles, $A$ is a set of attributes, $V$ is a set of product attribute values, and $O$ is a set of observed triples in the PG. Note that we have open-world assumption and thus cannot predetermine the possible values of $V$. Triples in PG are attribute triples defined in Definition \ref{def:triple}. 
Fig.~\ref{fig::PG} illustrates an example PG.


We can now formally define the problem of {\em error detection in PG} as follows:

\textbf{Given:} a product graph $\mathcal G = \{T, V, A, O \}$. 

\textbf{Identify:} incorrect triples $\{(t, a, v)\}\subset O$.


\section{Our Proposed Framework: PGE}\label{sec:PGE}
\begin{figure*}
\centering
\includegraphics[width=0.8\linewidth]{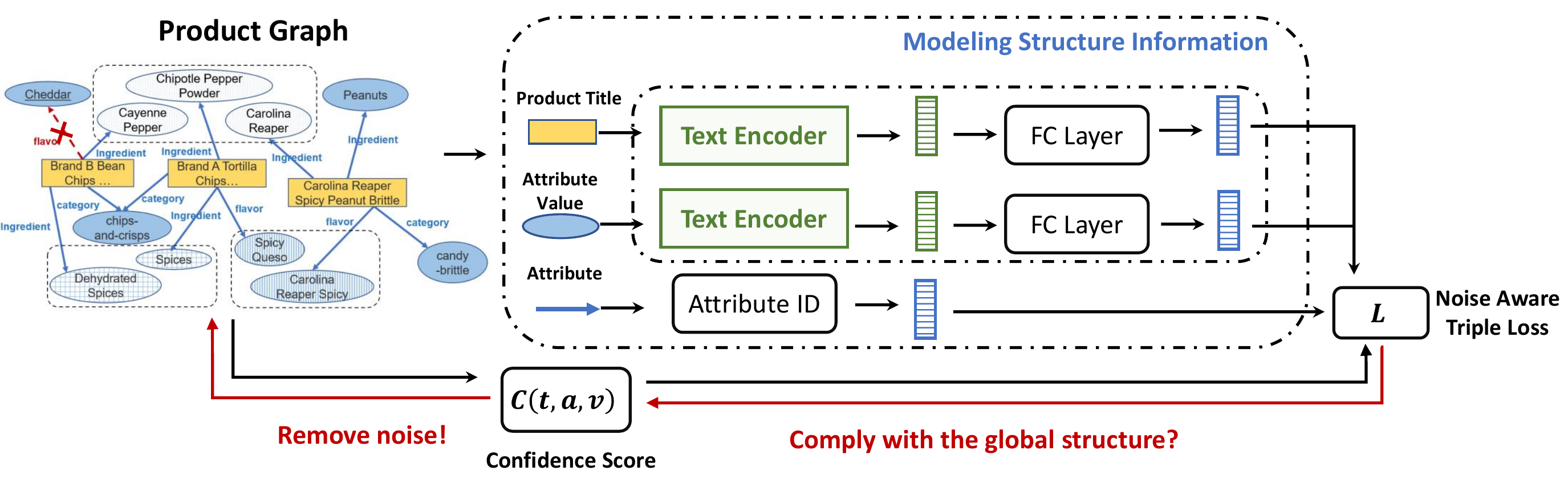}
\caption{\small{Illustration of the end-to-end PGE framework. The embedding vectors in green are text-based entities representations learned from text descriptions, while the embedding vectors in blue are the final entity embeddings learned under the guidance of the PG network structure. The arrows in red illustrate how the noise-aware mechanism removes noises in PG.}}\label{fig:framework}
\vspace{-0.1in}
\end{figure*}

In this section, we present PGE that learns the embeddings of PG entities by incorporating both the text information and the network structure of a PG to detect erroneous triples.
As shown in Fig.~\ref{fig:framework}, the framework includes three key components: (1) Learn text-based representations of entities from their raw text values; 
(2) Leverage network structure of a PG to guide the final embedding learning for error detection; 
(3) Introduce a noise-aware mechanism to 
diminish the impact of noisy triples to the representation learning.

\subsection{Text-based Representation Learning for Entities}\label{sec:cnn encoder}

\begin{figure}
\centering
\includegraphics[width=\linewidth]{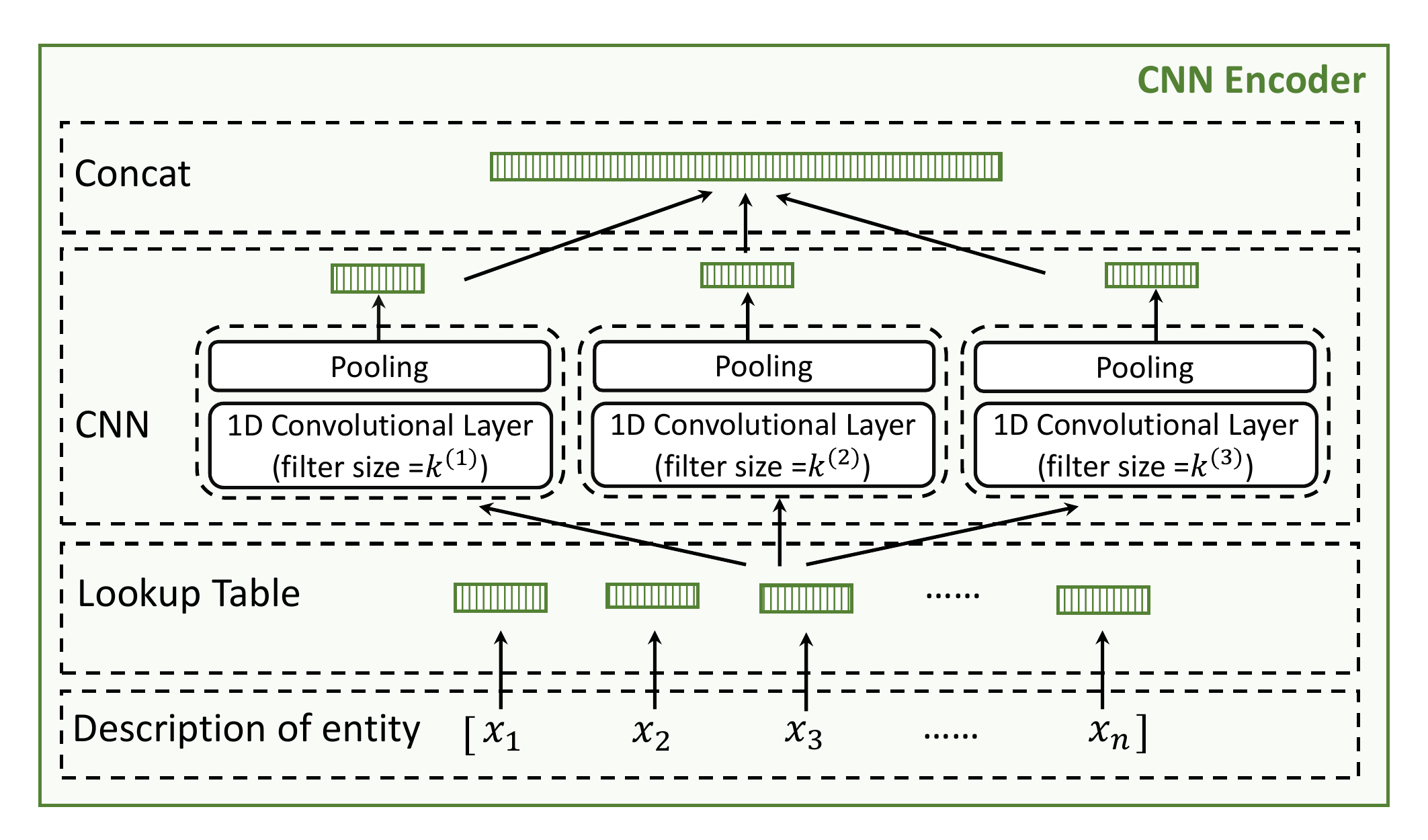}
\caption{\small{CNN-based text encoder.}}\label{fig:entity_encoder}
\vspace{-0.1in}
\end{figure}

In a typical KG embedding learning procedure, each entity is given a unique id which is then mapped to a learnable embedding. This approach is not optimal for PG embedding learning, because product titles ($T$) and attribute values ($V$) in a PG are mostly unstructured text containing rich semantic information, thus learning entity embeddings from only their ids
not only creates unnecessary degrees of freedom, but also discards their underlying semantic connections. For instance, the embeddings of product titles ``Brand A Tortilla Chips Spicy Queso,  6 - 2 oz bags'' and ``Brand B Bean Chips Spicy Queso, High Protein and Fiber, Gluten Free, Vegan Snack, 5.5 Ounce (Pack of 6)'' should be close to each other because they are semantically similar. There are several methods, such as convolutional neural network (CNN)-based methods and Transformer-based methods (e.g., BERT), that could be leveraged to learn the representations of product titles ($T$) and attribute values ($V$) in order to capture their semantic similarities. We present
scalability analysis of both text encoders in Section~\ref{sec:Scalability Analysis}. Due to the huge number of products contained in PGs, 
we pick the CNN architecture for its good scalability as well as effectiveness on many natural language processing tasks~\cite{rakhlin2016convolutional}. 
As shown in Figure~\ref{fig:entity_encoder}, the CNN encoder takes the raw text of a product title or an attribute value as the input and output its text-based representation. 
The first layer in the encoder transforms every word in the sequence into its respective embedding (initialized with word2vec~\cite{mikolov2013efficient}). 
The word embeddings then pass through three 1-d shallow CNNs with different filter sizes, which create three feature maps. Here we use different filter sizes to capture local semantic information from different text spans.
The final text-based representation of an entity 
is the concatenated feature maps learned by all CNNs. 

\subsection{Leverage Graph Structure to Guide Embedding Learning}\label{sec:KGE}
Manually labeled data are costly to obtain given the huge number of products in a PG. Fortunately, the rich structure information of a PG bridges the gap between the difficulties in obtaining labeled data and the necessity of supervision to detect errors. Although several recent papers have proposed to combine the text and structure information for KG representation learning, 
most of them ~\cite{xie2016representation,an2018accurate} learn two independent representations with separate loss functions and then integrate them with a linear combination. Such solution cannot generate a desired unified representation. 
To address this issue, we propose to learn the embeddings of entities and relations end to end, encoding the network structure that triples should obey on top of their text-based representations.

As shown in Fig.~\ref{fig:framework}, 
we introduce a fully-connected neural network layer to transform a text-based representation into its final representation to encode the network structure of a PG.
Boldfaced $\mathbf{t},\mathbf{a},\mathbf{v}$ denote the final embedding vector of product title $t$, attribute $a$, attribute value $v$, respectively. Since the number of attributes in a PG is small and well-defined comparing to titles and attribute values, we use randomly initialized learnable vectors to represent relations instead of CNN encoders.
To capture the network structure of PG, we define the objective function by maximizing the joint probability of the observed triples given the embeddings of both entities and relations. In particular, we assume all triples are conditionally independent given the corresponding embeddings. Then the joint distribution of all the triples is defined as:
\begin{equation}
\begin{aligned}
P(O) &= \prod_{(t,a,v)\in O} P\Big((t,a,v)|\{\mathbf{t}\},\{\mathbf{a}\},\{\mathbf{v}\}\Big).\\
\end{aligned}
\end{equation}

Since our goal is to detect the incorrect attribute value $v$ in a triple $(t,a,v)$, we optimize $P(v|t,a,\{\mathbf{t}\},\{\mathbf{a}\},\{\mathbf{v}\})$ instead of $P(t,a,v)|\{\mathbf{t}\},$ $\{\mathbf{a}\},\{\mathbf{v}\})$, which can be formalized as follows:

\begin{equation}
\begin{aligned}
P(v|t,a,\{\mathbf{t}\},\{\mathbf{a}\},\{\mathbf{v}\}) = \frac{\exp\Big(f_{a}(t,v)\Big) }{\sum_{v'\in V} \exp\Big(f_{a}(t,v')\Big)}
\end{aligned}
\end{equation}
where $f_{a}(t,v)$ can be defined by any KG embedding scoring functions. For example, in TransE, $f_{a}(t, v) = \gamma - \left \| \mathbf{t} + \mathbf{a} - \mathbf{v}\right \|_{1}^2$, where $\mathbf{t},\mathbf{a},\mathbf{v} \in \mathbb{R}^d$ and $\gamma$ is a fixed margin. In particular, a higher $f_{a}(t,v)$ usually indicates that the triple $(t,a,v)$ is more plausible. Due to the large number of attribute values $|V|$ involved in a PG, it is impractical to directly compute the softmax functions. Therefore, we adopt negative sampling~\cite{mikolov2013distributed} as computationally efficient approximation instead and reformulate the objective function as follows: 
\begin{equation}\label{eq:KGE}
\begin{aligned}
&\sum_{(t,a,v) \in O} \Big[-\log \sigma\Big(f_a(t,v)\Big) \\
-& \frac{1}{|\mathcal{N}(t, a, v)|} \sum_{(t, a, v^{\prime})\in \mathcal{N}(t, a, v)}\log \sigma\Big(-f_a(t,v^{\prime})\Big)\Big]
\end{aligned}
\end{equation}
where $\sigma$ is the standard sigmoid function, $O$ represents the observed facts in PG, $\mathcal{N}(t, a, v)$ is a set of negative samples for an attribute triple $(t,a,v)$. More specifically, for each observed triple $(t, a, v)$ we sample a set of negative samples $\mathcal{N}(t, a, v) \subset \{(t,a,v')|v' \in V\}$ by replacing the attribute value $v$ with a random value from $V$.

\subsection{Noise-aware Mechanism}\label{sec:noise aware}
The objective function in Eq. (\ref{eq:KGE}) indiscriminately minimizes the scores of all facts in PG without taking their trustworthiness into consideration. As a result, noisy facts can mislead the embedding model to learn wrong structure information, thus harm the performance of embeddings in error detection.

To address this issue, we propose 
a novel noise-aware mechanism to reduce the impact of noisy triples on the representation learning process. Knowledge representations are learned to ensure global consistency with all triples in PG. Correct triples are inherently consistent, which can jointly represent the global network structure of PG; noisy triples usually conflict with these global network structures. Consequently, by forcing consistency between correct triples and noises, performance is unnecessarily sacrificed. The main idea of the noise-aware mechanism is to explicitly allow the model to identify and ``markdown'' a small set of incorrect triples during training and reduce their impact on the loss function.

More specifically, we introduce a binary learnable confidence score, $C(t,a,v)$, for every triple $(t, a, v)$ in a PG to indicate whether the fact is true or false. $C(t,a,v) = 1$ indicates the triple is correct and $0$ otherwise. Associating confidence scores with triples in a PG actively downweight potential noises in the PG. The objective function of the noise-aware PGE model is defined as follows.

\begin{equation}
\label{eq:noise-aware-obj1}
\begin{aligned}
\mathcal{L} &= \sum_{(t,a,v) \in O}C(t,a,v) \Big[-\log \sigma\Big(f_a(t,v)\Big) \\
&-\frac{1}{|\mathcal{N}(t, a, v)|}\sum_{(t, a, v^{\prime})\in \mathcal{N}(t, a, v)}\log \sigma\Big(-f_a(t,v^{\prime})\Big)\Big] \\
&+ \alpha \sum_{(t,a,v) \in O} \Big(1-C(t,a,v)\Big) \\
& s.t., C(t,a,v) \in \{0,1\}
\end{aligned}
\end{equation}
where $C(t,a,v)$ is the binary confidence score assigned to a triple $(t,a,v)$ in a PG, and $\alpha\sum_{(t,a,v) \in O} \Big(1-C(t,a,v)\Big)$ is a regularization term imposed on confidence scores to control their sparsity. The problem in Eq.(~\ref{eq:noise-aware-obj1}) is difficult to solve due to the boolean constraint on $C(t,a,v)$. Following the common relaxation technique in ~\cite{von2007tutorial}, the boolean constraint on $C(t,a,v)$ can be relaxed as:
\begin{equation}\label{eq:relaxation}
\begin{aligned}
C(t,a,v)^2+\Big(1-C(t,a,v)\Big)^2 =1,
\end{aligned}
\end{equation}
since minimizing $1-C(t,a,v)^2-\Big(1-C(t,a,v)\Big)^2$ polarizes $C(t,a,v)$. Therefore, we rewrite the Eq.~(\ref{eq:noise-aware-obj1}) as:
\begin{equation}
\label{eq:noise-aware-obj2}
\begin{aligned}
\mathcal{L} &= \sum_{(t,a,v) \in O}C(t,a,v) \Big[-\log \sigma\Big(f_a(t,v)\Big) \\
&-\frac{1}{|\mathcal{N}(t, a, v)|}\sum_{(t, a, v^{\prime})\in \mathcal{N}(t, a, v)}\log \sigma\Big(-f_a(t,v^{\prime})\Big)\Big] \\
&+ \alpha \sum_{(t,a,v) \in O} \Big(1-C(t,a,v)\Big) \\
&+ \beta \sum_{(t,a,v) \in O}\Big(1- C(t,a,v)^2-\Big(1-C(t,a,v)\Big)^2 \Big). \\
\end{aligned}
\end{equation}

\section{Experiments}
\subsection{Dataset} 
We evaluate our PGE on two datasets: one real-world e-commerce dataset collected from publicly available Amazon webpages, and one widely used benchmark dataset FB15K-237. Table~\ref{Statistics} summarizes the statistics of both datasets. 

\textbf{Amazon Dataset:} To evaluate PGE on real-world e-commerce dataset, we construct a product graph with the product data obtained from public Amazon website. Each product in the Amazon dataset is associated with multiple attributes, such as product title, brand and flavor, whose values are short texts. As shown in Table~{\ref{Statistics}}, the Amazon dataset contains 750,000 products associated with 27 structured attributes and 5 million triples. To avoid bias, we sampled products from 325 product categories across different domains, such as food, beauty and drug. To prepare labeled test data, we asked Amazon Mechanical Turk (MTurk) workers to manually label the correctness of two attributes, including \textit{flavor} and \textit{scent}, based on corresponding product profiles. Each data point is annotated by three Amazon Mechanical Turk workers and the final label is decided by majority voting. Among 5,782 test triples, 2,930 are labeled as incorrect and 3,304 are labeled as correct.

\textbf{FB15K-237:} The FB15K dataset is the most commonly used benchmark knowledge graph dataset
~\cite{bordes2013translating}. It contains knowledge graph relation triples and textual mentions of Freebase entity pairs. FB15K-237 is a variant of FB15K dataset where inverse relations are removed to avoid information leakage problem in test dataset. The FB15K-237 datasets benefit from human curation that results in highly reliable facts. We add 10\% noisy triples to the data set by randomly sample 10\% triples and substituting the original head or tail entity with a randomly selected entity.

\begin{table*}
\centering
\begin{tabular}{cccccccc} 
  \toprule 
  Dataset & \#Relations & \#Entities & \#Products & \#Attributed values & \#Train & \#Valid & \#Test   \\ 
  \hline
  Amazon Dataset & 27 & 1,017,374 & 750,000 & 267,374 & 4,989,375 & 6,924 & 5,782\\
  \hline 
  FB15K-237 & 234 & 13,714 & - & - & 67,894 & 2,750 & 3,042 \\ 
  \bottomrule 
\end{tabular}
\caption{\small{Data statistics}}\label{Statistics}
\vspace{-0.1in}
\end{table*}

\subsection{Experimental Setting}
Our goal is to identify incorrect attribute values of a product, which can be formally defined as a triple classification problem in PG. 
We choose a threshold $\theta$ based on the best classification accuracies on the validation dataset, then classify a triple $(t, a, v)$ as correct if its score $f_a(t, v) > \theta$, otherwise incorrect. We apply the same settings to all baseline methods to ensure a fair comparison. We evaluate two versions of our model by incorporating TransE~\citep{bordes2013translating} and RotatE~\citep{sun2019rotate} as the score function, respectively. 
 
\textbf{Evaluation Metric.}
We adopt the area under the Precision-Recall curve (PR AUC) and Recall at Precision=X (R@P=X) to evaluate the performance of the models. To be more specific, PR AUC is defined as the area under the precision-recall curve, which is widely used to evaluate the ranked retrieval results. R@P is defined as the recall value at a given precision, which aims to evaluate the model performance when a specific precision requirement needs to be satisfied. For example, R@R = 0.7 shows the recall when the precision is 0.7.

\textbf{Compared Methods.} We evaluate PGE against state-of-the-art (SOTA) algorithms, including (1) NLP-based method (LSTM, Transformer~\cite{vaswani2017attention}); (2) structure based KG embedding (TransE~\cite{bordes2013translating}, DistMult~\cite{yang2014embedding}, ComplEx~\cite{trouillon2016complex}, RotatE~\cite{sun2019rotate}); (3) text and KG joint embedding (e.g., DKRL~\cite{xie2016representation}, SSP~\cite{xiao2017ssp}); and (4) noise-aware KG embedding (CKRL~\citep{xie2018does}). We choose CNN and BERT as the text encoders of PGE. Since BERT cannot handle Amazon dataset due to scalability issues, only the results of CNN is reported in Section~\ref{sec:Transductive Setting} and Section~\ref{sec:Inductive Setting}. We present scalability analysis of both text encoders in Section~\ref{sec:Scalability Analysis}.
We also include the approach ``Union of Transformer and PGE'' to show how PGE complement Transformer. To combine Transformer and PGE for error detection, the approach ``Union of Transformer and PGE'' re-ranks the test triples by jointly considering the ranking given by the Transformer and PGE. For example, given a test triple $(h,r,t)$, suppose Transformer rank it as i while PGE rank it as j. Then the average ranking of triple $(h,r,t)$ is $R_{avg}^{(h,r,t)} = (1/i+1/j)/2$. Based on $R_{avg}^{(h,r,t)}$, ``Union of Transformer and PGE'' re-ranks the test triples. Smaller $R_{avg}^{(h,r,t)}$ results in higher ranking assigned by ``Union of Transformer and PGE''.
In addition to ``Union of Transformer and PGE'', we also include a strong ensemble method - RotatE+ to enrich knowledge graph with information extraction technique. In particular, RotatE+ first applies OpenTag~\cite{zheng2018opentag, karamanolakis2020txtract}, the SOTA information extraction toolkit developed by Amazon Product Graph Team, to extract all relevant attributes from product title and product description to enrich the PG, then applies KG embedding method RotatE on the enriched KG to detect the error.


\textbf{Setup Details.} 
In data preprocessing, we remove all stop words from raw texts and map words to 300-dimensional word2vec vectors trained with GoogleNews. We adopt the Adam~\cite{kingma2014adam} optimizer with learning rate among $\{0.0001, 0.0002, 0.0005\}$ following~\cite{sun2019rotate}, and margin $\gamma$ among $\{12.0, 24.0\}$. For the CNN encoder, we try different filter sizes among $\{1,2,3,4\}$ for different CNNs. To fairly compare with different baseline methods, we set the parameters for all baseline methods by a grid search strategy. The best results of baseline methods are used to compare with PGE. 

\subsection{Transductive Setting}\label{sec:Transductive Setting}
Transductive setting focuses on the situation where all attribute values in the test triples have been observed in the training stage. To compare different algorithms on the triple classification task, we require each method to predict the correctness of triples in the test dataset. Table~\ref{tab::transductive} shows the comparison results. Here are several interesting observations: (1) PGE consistently outperforms KG embedding models as well as CKRL in all cases with significant performance gain (improving by 24\% - 30\%  on PR AUC), 
which ascribes to the utilization of textual information associated with entities; (2) PGE also obtains better performance than NLP-based approaches as they cannot leverage graph structure information in KGs. In particular, NLP-based methods show the worst performance on the FB15k-237 dataset while the second best performance on Amazon dataset. The major reason is that FB15k-237 contains much richer graph information 
compared to the Amazon dataset (i.e., there are 27 attributes in Amazon dataset while 234 relations in FB15k-237). Therefore, graph structure plays a more critical role in
error detection task in FB15k-237; (3) PGE shows better performance compared to DKRL and SSP. The major reason is that DKRL and SSP learn the structural representations and the textual representations by separate functions.

\begin{table*}[t]
\centering
\resizebox{\textwidth}{!}{
\begin{threeparttable} 
    \begin{tabular}{l|c|ccccc|ccccc}
    \hline
    \multirow{2}{*}{Categories} & \multirow{2}{*}{Method} & \multicolumn{5}{c|}{Amazon Dataset} & \multicolumn{5}{c}{FB15k-237} \\ 
    & & PR AUC & R@P=0.7 & R@P=0.8 & R@P=0.9 &Time (hours) &PR AUC & R@P=0.7 & R@P=0.8 & R@P=0.9 &Time (hours) \\
    \hline
    \multirow{2}{*}{NLP-based methods} & LSTM  & 0.704 & 0.572 & 0.416 & 0.159 & 16.32 & 0.626& 0.595 & 0.445 & 0.239 & 0.43\\
    & Transformer~\cite{vaswani2017attention} & 0.719 & 0.601 & 0.427 & 0.194 & 79.46 & 0.648 & 0.649 & 0.503 & 0.245 & 12.82 \\
    \hline
    \multirow{4}{*}{Structured based KG embedding} & TransE~\cite{bordes2013translating} & 0.584 & 0.390 & 0.308 & 0.213 & 20.57 & 0.772 & 0.793 & 0.737 & 0.685 & 0.58 \\
    & DistMult~\cite{yang2014embedding} & 0.573  & 0.362 & 0.291 & 0.197 & 32.86 &
    0.819 & 0.872 & 0.813 & 0.751 & 4.12 \\
    & ComplEx~\cite{trouillon2016complex} & 0.579 & 0.373 & 0.310 & 0.207 & 36.31 & 0.781 & 0.814 & 0.759 & 0.712 & 5.16 \\
    & RotatE~\cite{sun2019rotate} & 0.597 & 0.405 & 0.336 & 0.239 & 35.11 & 0.824 & 0.875  & 0.823 & 0.766 & 5.33\\
    & RotatE+\tnote{$\ddag$} & 0.611  & 0.423 & 0.369 & 0.221 & 36.79 & - & - & - & - & -\\
    \hline
    \multirow{2}{*}{Text and KG joint embedding} & DKRL~\cite{xie2016representation} & 0.693 & 0.552 & 0.408 & 0.246 & 45.38 & 0.909 & 0.945 & 0.901 & 0.868 &7.25  \\
  & SSP~\cite{xiao2017ssp}\tnote{$\dag$} & - & - & - & - & - & 0.927 & 0.951 & 0.915 & 0.882 & -  \\
    \hline
    Noise-aware KG embedding & CKRL~\citep{xie2018does} & 0.586 & 0.392 & 0.304 & 0.217 & 21.16 & 0.768 & 0.725 & 0.672 & 0.627 & 0.62  \\
    \hline
    \multirow{2}{*}{Our Proposed model} & PGE(CNN)-TransE & 0.738 & 0.690 & 0.436 & 0.267 & 23.12 & \textbf{0.990} & \textbf{0.997} & \textbf{0.995} & \textbf{0.986} & 0.67 \\
    & PGE(CNN)-RotatE & \underline{0.745} & \underline{0.729} & \textbf{0.516} & \underline{0.325} & 39.41 & \textbf{0.990} & \textbf{0.997} & \underline{0.993} & \underline{0.983} & 5.71\\
    \hline
    \multicolumn{2}{c|}{Union of Transformer and PGE(CNN)-RotatE} & \textbf{0.751} & \textbf{0.747} & \underline{0.509} & \textbf{0.349} & - & \underline{0.938} & \underline{0.958} & 0.911 & 0.893 & -\\
    \hline
    \end{tabular}
\begin{tablenotes} 
\footnotesize
\item[$\dag$] Since SSP cannot handle Amazon dataset due to scalability issues, only the results on the FB15K-237 is reported.
\item[$\ddag$] RotatE+ first applies OpenTag~\cite{zheng2018opentag, karamanolakis2020txtract}, an information extraction toolkit developed by Amazon Product Graph Team, to extract all relevant attributes from product title and product description to enrich the product graph, then apply RotatE~\cite{sun2019rotate} on the enriched KG to detect the error.
\end{tablenotes} 
\end{threeparttable}
}
\caption{\small{Results of error detection under the transductive setting. The numbers in bold represent the best performance among all methods while the numbers underlined represent the second best. Evaluation of PGE on the Amazon dataset shows that PGE is able to improve over the SOTA methods on average by 18\% on PR AUC.}}\label{tab::transductive}
\vspace{-0.1in}
\end{table*}

\subsection{Inductive Setting} \label{sec:Inductive Setting}
Inductive setting focuses on the situation where attribute values in the test triples are not presented in a PG, which is a common scenario for PG error detection. Existing KG embedding models 
are not effective in dealing with this situation because they cannot generate representations for the entities outside of KGs due to missing ids. 
Therefore, we do not include them as baselines in this subsection. Unlike the KG embedding methods, which map entities to their embeddings using ids, our proposed PGE learns embeddings of entities based on their text-based representations and thus can naturally handle the inductive setting. 

To prepare an inductive setting, we filter the training set by excluding any triples that share entities with the selected test triples, so that the training and the testing use disjoint sets of entities. We report the results on R@P=0.6, R@P=0.7, R@P=0.8 in Table~\ref{tab::comparision}. We observe that: (1) All methods perform worse in the inductive setting without exception, which indicates that inductive setting is indeed more challenging; (2) NLP-based methods perform the best among all methods. The major reason is that language naturally has strong transferring ability while PGE still relies on the graph structure to make the prediction. Although text encode can transferring information among entities, it doesn't help to predict a never seen graph structure; (3) Although NLP-based methods perform better than PGE on the Amazon dataset, the best results are given by the union of Transformer and PGE (improving by 9\% on R@P=0.9 compared with Transformer), showing that PGE can learn the undetected error by Transformer
; (4) Although PGE cannot leverage textual information as well as Transformer (because CNN is less powerful compared with Transformer in capturing the semantic information. Not to mention we employ shallow CNN as text encode due to scalability issues), it still achieves comparable result on the Amazon dataset. Moreover, they achieve the SOTA on FB15k-237, which further validates the strong ability of PGE in detecting errors in a KG with rich textual information; (5) DKRL and SSP perform the worst among all methods, which again demonstrates their weakness.
\begin{table*}[t]
\centering
\resizebox{\textwidth}{!}{
\begin{threeparttable} 
    \begin{tabular}{l|c|cccc|cccc}
    \hline
    \multirow{2}{*}{Categories} & \multirow{2}{*}{Method} & \multicolumn{4}{c|}{Amazon Dataset} & \multicolumn{4}{c}{FB15k-237} \\ 
    & &PR AUC & R@P=0.6 & R@P=0.7 & R@P=0.8 &PR AUC & R@P=0.6 & R@P=0.7 & R@P=0.8 \\
    \hline
    \multirow{2}{*}{NLP-based methods} & LSTM  & 0.626 & 0.756 & 0.476 & 0.340 & 0.581 & 0.717 & 0.436 & 0.204 \\
    & Transformer~\cite{vaswani2017attention} & \underline{0.643}& \underline{0.771} & \underline{0.495} & \underline{0.354}  & 0.603  & 0.748 & 0.453 & 0.238  \\
    \hline
    \multirow{2}{*}{Text and KG joint embedding} & DKRL~\cite{xie2016representation} & 0.552 & 0.593 & 0.252 & 0.068 & 0.698 & 0.790 & 0.638 & 0.415 \\
 & SSP~\cite{xiao2017ssp}\tnote{$\dag$} & - & - & - & - & 0.716 & 0.807 & 0.654 & 0.419 \\
    \hline
    \multirow{2}{*}{Our Proposed model} & PGE(CNN)-TransE & 0.585 & 0.730 & 0.412 & 0.197 & 0.787 & 0.871 & 0.724 & 0.674 \\
    & PGE(CNN)-RotatE & 0.596 & 0.741 & 0.437 & 0.228 & \textbf{0.836} & \underline{0.919} & \textbf{0.845} & \textbf{0.753} \\
    \hline
    \multicolumn{2}{c|}{Union of Transformer and PGE(CNN)-RotatE} & \textbf{0.649} & \textbf{0.779} & \textbf{0.512} & \textbf{0.386}  & \underline{0.833} & \textbf{0.923} & \underline{0.837} & \underline{0.743} \\
    \hline
    \end{tabular}
\begin{tablenotes} 
\footnotesize
\item[$\dag$] Since SSP cannot handle Amazon dataset due to scalability issues, only the results on the FB15K-237 is reported.
\end{tablenotes} 
\end{threeparttable}
}
\caption{\small{Results of error detection under the inductive setting. The bold numbers represent the best performances among all methods while the underlined numbers represent the second best. We observe that PGE achieves the SOTA on the structure-rich FB15k-237 data set. The best results on the Amazon dataset are given by the union of the Transformer model and PGE, showing that although PGE does not perform as well as NLP-based methods on the Amazon dataset, it complements Transformer for its strong ability in capturing graph structure.}}\label{tab::comparision}
\vspace{-0.1in}
\end{table*}

\subsection{Validity of Noise-aware Mechanism}
\begin{figure*}
\centering
\begin{minipage}[t]{0.65\linewidth}
\centering
\includegraphics[width=0.49\linewidth]{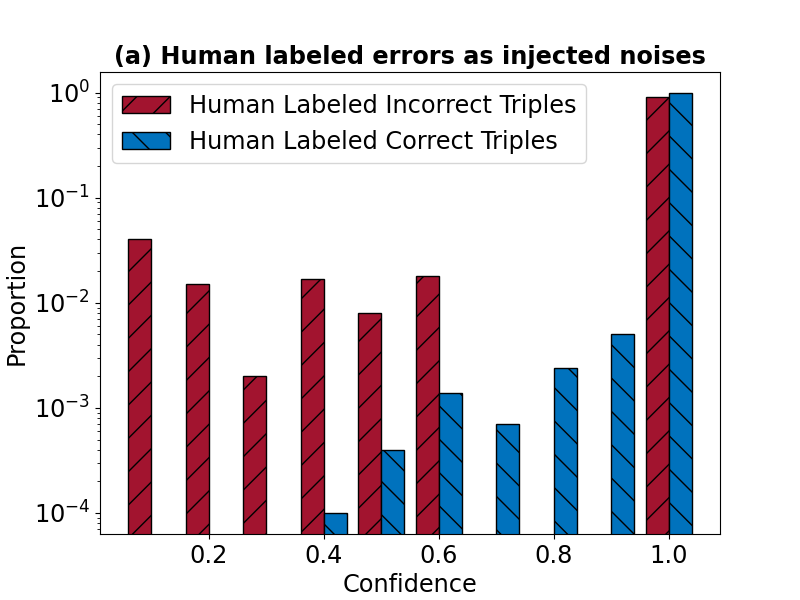}
\hfill
\includegraphics[width=0.49\linewidth]{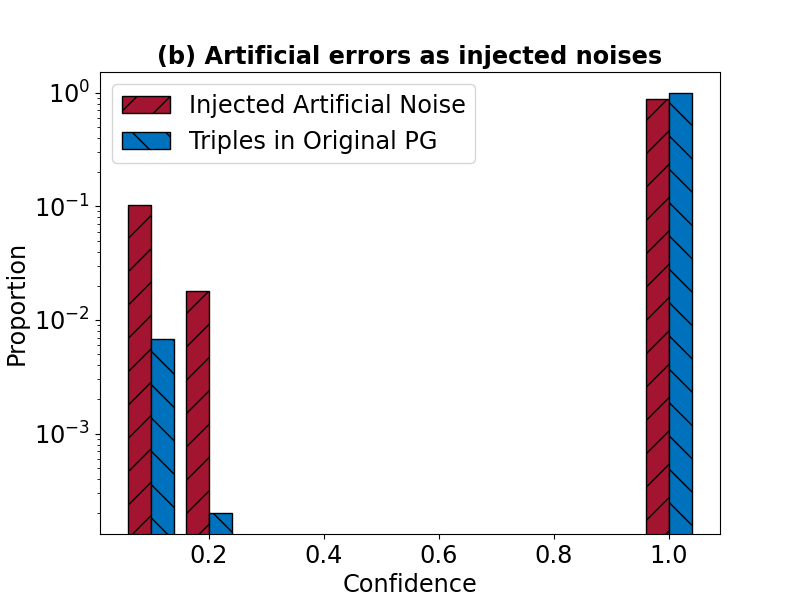}
\caption{\small{Distribution of confidence scores learned by PGE(CNN)-RotatE on the Amazon dataset with different injected noises.}}
\label{Fig.distribution_rotatE}
\end{minipage}
\hfill
\begin{minipage}[t]{0.32\linewidth}
\centering
\includegraphics[width=\linewidth]{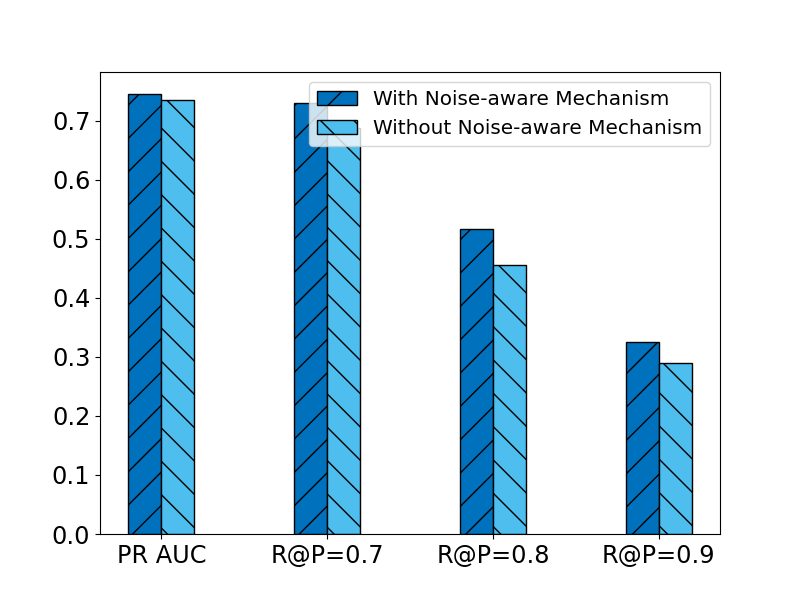}
\caption{\small{PGE(CNN)-RotatE with v.s. without noise-aware mechanism on noisy Amazon dataset.}}
\label{Fig.Noise Aware Mechanism_rotatE}
\end{minipage}
\end{figure*}

\textbf{Validity of Confidence Scores with Different Injected Noises.}
To evaluate the benefit of including confidence scores $C(t,a,v)$ in the noise-aware mechanism, shown in Eq.~(\mbox{\ref{eq:noise-aware-obj2}})
, we evaluate \emph{PGE(CNN)-RotatE} on the Amazon dataset with two different kinds of injected noises. First, we inject human-labeled correct triples and incorrect triples into the training data. Confidence scores are learned to determine the correctness of these injected labeled triples. The distribution of confidence scores are shown in Fig.~{\ref{Fig.distribution_rotatE}} (a). Second, we inject artificial noises into the training data.
We substitute attribute values of existing triples in the Amazon dataset with a random value to generate these artificial noises.
Confidence scores are learned to distinguish artificial noises from the original triples. Fig.~{\ref{Fig.distribution_rotatE}} (b) shows the distribution of confidence scores. The red bars represent the distribution of confidence scores for human-labeled incorrect triples (or injected artificial noises) while the blue bars represent the distribution of confidence scores for human-labeled correct triples (or triples in the original Amazon dataset). 
We observe that real-world noises are more difficult to identify compared to artificial noises. Despite the difficulty in detecting the real-world error, confidence scores of human-labeled correct triples are mainly over $0.5$, validating the promising capability of the confidence scores to distinguish noises in PG. In addition, we observe that 1\% triples in the original Amazon dataset have also been identified as noises in Fig.~{\ref{Fig.distribution_rotatE}} (b). We have verified that most of these triples are indeed noisy triples in the original Amazon dataset.

\textbf{Overall Impact of Noise-aware Mechanism.}
To further validate the overall benefits brought by noise-aware mechanism, we also evaluate \emph{PGE(CNN)-RotatE} without noise-aware mechanism on Amazon dataset used in Section~{\ref{sec:Transductive Setting}}. Figure~{\ref{Fig.Noise Aware Mechanism_rotatE}} presents the comparison results. We observe that the noise-aware mechanism brings significant performance gain: PGE(CNN)-RotatE with noise-aware mechanism increases the PR AUC of PGE(CNN)-RotatE without noise-aware mechanism from $0.734$ to $0.747$ and increases R@P=0.9 from $0.289$ to $0.325$. 

\subsection{Scalability Analysis}~\label{sec:Scalability Analysis}
To demonstrate the scalability of PGE, we present the training time of PGE on Amazon dataset of different sizes in Table~{\ref{tab:training time}}. We vary the sample ratio among $\{0.1, 0.3, 0.5, 0.7, 1\}$ to select only a portion of triples in Amazon dataset to construct PG of different sizes. Two text encoders, CNN-based text encoder and BERT-based text encoder, are leveraged to learn entity representations. In particular, BERT-based text encoder takes the raw text of product titles or attribute values as input. The first token of input is always a special classification token ([CLS]). The final hidden state corresponding to this token is used as the text-based representation of entities. We observe that \emph{PGE(BERT)-RotatE} cannot be applied to Amazon dataset due to the scalability issue. It takes near 2 days for 10\% data and over 3 days for 30\% data. Therefore, we focus on CNN in this paper. We observe that \emph{PGE(CNN)-RotatE} scales up to large datasets with similar scalability compared to KGE model.
\begin{table}
\centering
\resizebox{\linewidth}{!}{
\begin{tabular}{cccccc} 
\toprule 
\multirow{2}{*}{Models}   & \multicolumn{5}{c}{Percentage of Sampled Triples} \\ 
\cline{2-6}
   & 0.1 & 0.3 & 0.5 & 0.7 & 1 \\ 
\hline


RotatE & 3.22 & 10.77 & 17.63 & 24.86  & 35.11 \\

PGE(CNN)-RotatE & 4.07 & 11.95 & 19.44  & 27.62 & 39.41 \\ 

PGE(BERT)-RotatE & 45.21 & > 3 day & > 3 day & > 3 day & > 3 day \\
  
  \bottomrule 
\end{tabular}
}
\caption{\small{Training Time (hours) of different methods on Amazon dataset.}}\label{tab:training time}
\vspace{-0.2in}
\end{table}

\subsection{Case Study}
Previous experiments have shown the promising performance of PGE in both transductive setting and inductive setting. To further demonstrate the capability of PGE in detecting real-world errors in PG, we conduct case study to give examples of identified errors in the Amazon dataset as shown in Table~\ref{case study of noise}. We use PGE(CNN)-RotatE to evaluate if a triple is a correct fact. Threshold $\sigma$ is chosen based on the best classification accuracies on the validation dataset in order to classify triples.
We observe that most attribute values of identified errors violate the global graph structure of PG and thus can be classified as errors. For example, product 2,3, and 4 in Table~\ref{case study of noise} are not groceries thus should not have the attribute ``flavor''. Although the attribute values of product 1 and 5 include commonly observed phrases to describe the scent, the word ``conditioner'' in the attribute values makes them no longer correct attribute values of ``scent''. This observation shows that PGE not only leverages the graph structure of PG to detect noise (e.g., example 2,3,4) but also show sensitivity to subtle differences of language. 



\begin{table}
\centering
\resizebox{\linewidth}{!}{
\begin{threeparttable} 
 \begin{tabular}{cccc} 
  \toprule 
Product & Attribute & Attribute Value  \\ 
\midrule 
  \tabincell{c}{Pure Mint Shampoo and \\ Hair Conditioner for Women \\ and Men - 10 oz}& scent & \tabincell{c}{mint shampoo \\ and conditioner set} \\
 \midrule 
  \tabincell{c}{Brand A Foot Brush \\and Pumice (Pack of 4)}\tnote{$\dag$} & flavor & bamboo \\
  \midrule 
  \tabincell{c}{Brand B Sweet BBQ Rub 11.2 oz}\tnote{$\dag$} & flavor & sweet  \\
  \midrule 
  \tabincell{c}{Hassle Free Storage Pop-Up \\ Mesh Laundry Hamper (Aqua) }& flavor &  octopus \\
  \midrule 
  \tabincell{c}{Brand C Organics Conditioner, \\ Tea Tree Oil \& Blue Cypress, \\ 12 Ounce (Pack of 3)}\tnote{$\dag$} & scent & \tabincell{c}{conditioner \\ tea tree oil \\ and blue cypress} \\
  \bottomrule 
 \end{tabular} 
\begin{tablenotes} 
\footnotesize
\item[$\dag$] We mask the brand of the products to avoid revealing sensitive information.
\end{tablenotes} 
\end{threeparttable}
}
\caption{\small{Identified errors on Amazon dataset.}}
 \label{case study of noise}
 \vspace{-0.1in}
\end{table}

\section{Related Work}\label{sec:related_work} 
\textbf{Error Detection in Knowledge Graph.}
Most KG noise detection process is carried out when constructing KGs, such as Freebase, Google Knowledge Graph, Walmart product graph, YAGO, NELL, and Wikipedia~\cite{auer2007dbpedia, bollacker2008freebase, suchanek2007yago, heindorf2016vandalism,paulheim2017knowledge}. Despite the efforts during KG constructions, errors are widely observed in existing KGs. A recent open IE model on the benchmark achieves only 24\% precision when the recall is 67\%~\cite{stanovsky2018supervised} and the estimated precision of NELL is only 74\%~\cite{carlson2010toward}. 
To detect errors for an existing KG, most existing methods explore additional rules~\cite{beskales2010sampling, beskales2013relative, khayyat2015bigdansing,bohannon2007conditional,fan2008conditional,chu2013holistic,geerts2013llunatic,heidari2019holodetect,cortes2012semantics,fan2016functional}. 
Considering all kinds of errors that could be made in the real world, it is unrealistic to identify all required rules to cover all possible cases.
In contrast, our proposed method employs KG embedding model to automatically learn the correlation of entities, which could be considered as fuzzy rules to guide value cleaning in KGs. More recently, detecting noises while learning knowledge representations simultaneously becomes a hot topic. A confidence-aware framework CKRL~\citep{xie2018does} is proposed to incorporate triple confidence into KG embedding models to learn noise-aware KG representations. However, the confidence of triples are easily affected by model bias (i.e., improper order of triples in training sets may even amplify the impact of noises). In addition, it ignores the rich semantic information in KGs, which is strong evidence to judge triple quality. In this paper, we propose a noise-aware KG embedding learning method, which can utilize rich semantic information to identify noises, which conflict with the global network structures. 

\textbf{Knowledge Graph Embedding.}
Knowledge Graph Embedding (KGE) aims to capture the similarity of entities by projecting entities and relations into continuous low-dimensional vectors. Scoring functions, which measure the plausibility of triples in KGs, are the crux of KGE models. Representative KGE algorithms include TransE~\cite{bordes2013translating}, 
TransH~\cite{wang2014knowledge}, TransR~\cite{lin2015learning},
DistMult~\cite{yang2014embedding}, ComplEx~\cite{trouillon2016complex}, SimplE~\cite{kazemi2018simple} and RotatE~\cite{sun2019rotate}, which differ from each other with different scoring functions.

\textbf{Text and Knowledge Graph Joint Embeddings.}
In recent years, several attempts have been made to improve the knowledge representation by exploiting entity descriptions as additional information~\cite{an2018accurate,xu2016knowledge,xiao2017ssp}. However, the combination of the structural and textual representations is not well studied in these methods, in which two representations are learned by separate loss function or aligned only on the word-level.
As one of the most representative works, DKRL~\cite{xie2016representation} separates the objective function into two energy functions (i.e. one for structure and one for description) and integrates these two representations into a joint one by a linear combination.
Works proposed in~\cite{wang2014knowledge} and \cite{zhong2015aligning} align the entity name with its Wikipedia anchor on word level, which may lose some semantic information on the phrase or sentence level. 
SSP~\cite{xiao2017ssp} 
requires the topic model to learn pre-trained semantic vector of entities separately. 
Due to the rapid growth of pre-trained language representation models (PLM), several works are proposed to encode textual entity descriptions with a PLM as their embeddings. For example, KEPLER~\cite{wang1911kepler} proposes to encode textual entity descriptions with BERT as their embeddings, and then jointly optimize the KGE and language modeling objectives. BLP~\cite{daza2021inductive} trains PLM and KG in an end-to-end manner. Since the language modeling objective of PLM suffer from high computational cost and require a large corpus for training, it is time consuming to apply these methods to large scale KGs. In this paper, we propose an end-to-end method to jointly leverage both text information and graph structure for KG embedding learning in an efficient way.

\section{Conclusion}
In this paper, we propose a novel end-to-end noise-aware embedding learning framework, PGE, to learn embeddings on top of text-based representations of entities for error detection in PG. 
Experiment results on a real-world product graph show that PGE improves over state-of-the-art methods on average by 18\% on PR AUC in transductive setting. Although this paper focuses on the product domain, we also show in our experiments that,
the same techniques excel in other domains with textual information and noises. As the next step, we would investigate more efficient Transformer architecture to improve text encoder strength and efficiency of PGE. 
BERT-based text encoder is difficult to scale to large KG due to its full attention mechanism. To reduce the computation complexity of BERT-based text encoder, we can to extend the ideas of ~\cite{beltagy2020longformer, zaheer2020big} to allow sparse self-attention to tokens.
In addition, we can leverage additional information to improve the learned entity representations. For example, we could better capture the similarity among products by leveraging the hierarchical structure of product data or by leveraging the user behavior data.

\bibliographystyle{ACM-Reference-Format}
\bibliography{ref}

\end{document}